\documentclass[aps,pre,longbibliography,twocolumn]{revtex4-1}
\usepackage[utf8]{inputenc}

\usepackage{xcolor}
\usepackage{amsmath}
\usepackage{amsfonts}
\usepackage{amssymb}
\usepackage{makeidx}
\usepackage{graphicx}
\usepackage{tikz}
\usepackage{float}

\usepackage{lipsum}
\usepackage{circuitikz}

\usepackage[caption=false]{subfig}
\usepackage{siunitx}

\usepackage{color}
\definecolor{myblue}{RGB}{65,105,225}
\definecolor{mygreen}{RGB}{34,139,34}
\definecolor{myorange}{RGB}{255,69,0}
\usepackage[colorlinks,linkcolor=myorange,urlcolor=myblue,citecolor=myblue]{hyperref}

\newcommand{\tauc}{\tau_\text{c}}

\newcommand{\taud}{\tau_\text{d}}

\newcommand{\Lled}{B_{\mathrm{L}}}
\newcommand{\Lledmax}{B_{\mathrm{L}, \text{max}}}

\newcommand{\Vmax}{V_{\mathrm{C}, \text{max}}}

\begin{document}

\title{Measuring capacitor charge and discharge using an LED and a smartphone}

\author{R. Hurtado-Guti\'errez}
\email[]{rhurtado@onsager.ugr.es}
\affiliation{Departamento de Electromagnetismo y F\'{\i}sica de la Materia}
\affiliation{Institute Carlos I for Theoretical and Computational Physics, 18071 Granada, Spain}
\thanks{Both authors contributed equally}
\author{Á. Tejero}
\email[]{atejero@onsager.ugr.es}
\affiliation{Departamento de Electromagnetismo y F\'{\i}sica de la Materia}
\affiliation{Institute Carlos I for Theoretical and Computational Physics, 18071 Granada, Spain}
\thanks{Both authors contributed equally}

\begin{abstract}

In this article, we present a simple, inexpensive, and effective method for measuring the capacitor charge and discharge processes using a Light Emitting Diode (LED) and the light meter of a smartphone.
We propose a simple circuit in which the LED's brightness is linear on the capacitor's voltage, allowing us to use the smartphone to monitor the capacitor state accurately.
The method is tested experimentally, giving highly satisfactory results.
Its exceptional combination of accuracy, minimal requirements, and ease of setup makes it an excellent way to introduce undergraduate students to the concepts of electricity and electronics in any educational setting.

\end{abstract}
\maketitle

\section{Introduction and motivation}

In the recent years, a plethora of scientific articles related to the creation of virtual and homemade laboratories have appeared \cite{9951268, Monteiro_2023, haldolaarachchige2020lab, haldolaarachchige2020introductory, haldolaarachchige2021set, Raman_Achuthan_Nair_Nedungadi_2022, https://doi.org/10.1002/cae.20536, Gunawan_2018}.
This surge in interest has been especially motivated by the COVID-19 pandemic, during which students may not have had access to traditional laboratory equipment or may not have been able to attend in-person classes \cite{casaburo2021teaching, Franco_2023, Bjurholt_2023,Guo_2020}.
Particularly, in the field of physics, there were already existing specific web pages and resources. Mere outstanding examples include the University of Colorado's PhET \cite{phet} or the University of St Andrews' QuVis \cite{quvis}.
These webpages offer simulations to work with, where hands-on sessions can be designed for students of different levels. However, even though these virtual experiments provide invaluable insight, a crucial step in physics learning is related to manipulating real equipment and components, including a genuine visualization of the physical laws learned in theory.  In electronics education, Arduino \cite{arduino} has partially solved this issue \cite{bouquet2017project, haugen2014model, galeriu2015arduino, Erol_2023}. Its main benefit lies in the versatility it does offer, which allows one to build and integrate a wide range of sensors and automation systems. 
This makes it an excellent platform for introducing students to the basics of experimental physics, electronics and programming, allowing them to engage in hands-on experiments.
However, they are not always available in education centers and they require specific knowledge in the aforementioned areas.

Nevertheless, there is another versatile platform equipped with a vast selection of sensors that nearly everyone possesses in their pockets today: the smartphone.
Thanks to its wide availability and ease of use, the role of this device in education has been continuously increasing in the last few years \cite{kuhn2022smartphones, JIPF4167, gillen2022magnetic, gastaldi2020electronic, organtini2021physics}, especially due to the creation of specific software which allows the manipulation of their sensors \cite{phyphox, physicstoolbox}.
They make possible the design of cheap and intuitive experiments with everyday devices, which can be easily replicated by students at home and at low cost. This helps to engage students and make learning physics more accessible. 

\begin{figure}[t]

    \includegraphics[width=0.8\linewidth]{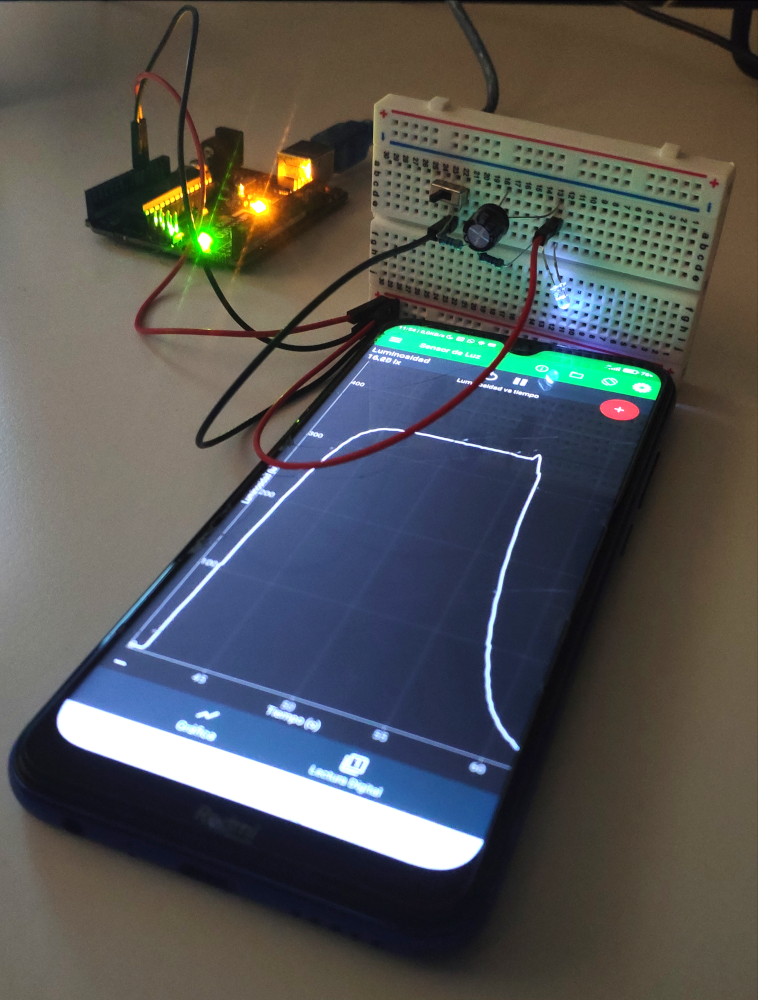}
    \caption{Experimental setup to measure the voltage of the capacitor through the LED brightness in both the charging and discharging processes. The breadboard is rotated and placed vertically so that the LED is pointing to the smartphone's light meter. We used a \emph{Xiaomi Redmi Note 8T} smartphone to test the method.}
    \label{fig:medidaexp}
\end{figure}

In line with this approach, we propose a novel method to study electric circuits using the light meter of a smartphone.
In particular, we consider a simple $RC$ circuit \cite{young2020university, Suits_2020} where the charging and discharging can be measured through the brightness of a LED, as seen in Fig. \ref{fig:medidaexp}.
As we will show, this not only constitutes an accessible, intuitive, and inexpensive way to illustrate the behavior of this circuit but also a surprisingly accurate method to measure it.
The article is structured as follows. First, we provide the relevant physical concepts involved in the experiment, where the circuit is presented and its dynamics are solved. Next, we present the experimental setup and the smartphone app used to measure the LED brightness. Then, we present the results of an actual realization of the experiment, confirming its validity. Finally, we explore an alternative arrangement of the LEDs to analyze the same phenomenon.

\section{Circuit Proposal and Solution}\label{sec:physics}

\begin{figure*}[htb]
    \subfloat{
        \begin{circuitikz}[american voltages]
            \ctikzset{bipoles/cuteswitch/thickness=0.25}
            \draw
             (-1,0) [cosw, -] to (0.4,0)
              to [R, l_=$R_1$] (2,0) 
              to [short, -] (2,0.7) 
              to [R, l_=$R_2$] (4,0.7) 
              to [empty led, l_=$V_\mathrm{L}$] (4.5,0.7) 
              to [short, -] (5,0.7) 
              to [short, -] (5,0) 
            
              (2,0) to [short, -] (2,-0.7) 
              to [C, l_=$C$] (5,-0.7) 
              to [short, -] (5,0)
              to [short, -] (5.5,0)
              to [short, -] (5.5,-2.5)
              to [V, l_=$V_\mathrm{S}$, invert] (-1,-2.5)
              to [short, -] (-1,0)

              (-1, -2.5) to [short,f_=$I$] ++(0,2.5)
              (3.7,-0.7) to [short,f_=$I_\text{C}$] (5,-0.7);
             \path
              (2,0.8) to [short, f=$I_\text{L}$] (4,0.8) 
              ;
        \end{circuitikz}

    }
    \hspace{2cm}%
    \subfloat{
        \includegraphics[width=0.45\linewidth]{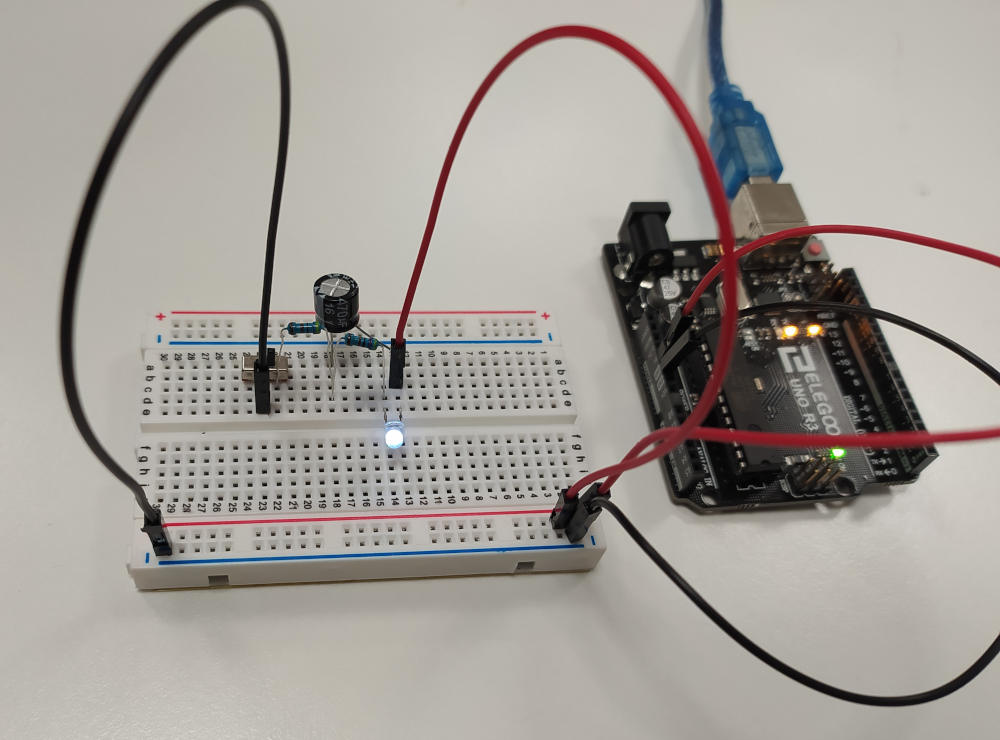}

    }
    \caption{\textbf{Left.} Diagram of the proposed circuit. \textbf{Right.} Setup on a breadboard with an Arduino-like board as a voltage source.}
    
    \label{fig:schemes}
\end{figure*}

    The circuit of interest in this article is shown in Fig.~\ref{fig:schemes}~(left). It is made of just a DC voltage source, a switch, two resistors, a capacitor, and an LED.
    The LED and one of the resistors are connected in parallel with the capacitor, so that the current across them (and therefore its brightness) is proportional to the capacitor voltage minus the LED forward voltage.
    This allows one to directly measure the state of the capacitor through the brightness of the LED.

    In what follows, we will first solve the equations of the circuit and then we will derive the relations between the capacitor voltage and the LED's brightness.
    The theoretical solution of the capacitor voltage in this circuit is illustrated in Fig.~\ref{fig:odes}.

\subsection{Charging circuit}
    \begin{figure}[b] 
        \centering
        \includegraphics[width=1\linewidth]{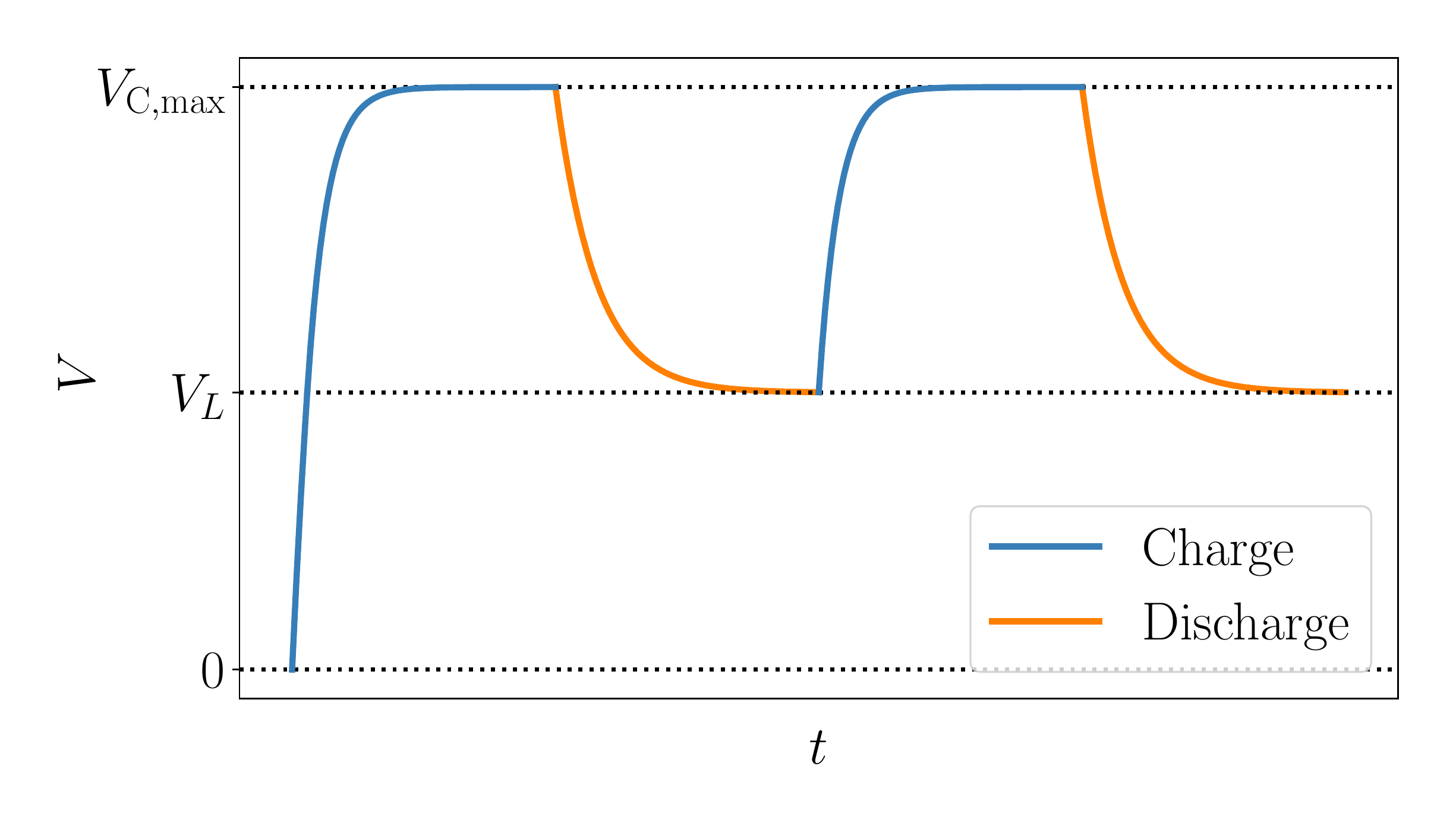}
    \caption{Theoretical solutions of the ODEs presented in Section \ref{sec:physics} for several charging and discharging cycles. At the beginning the tension in the capacitor is zero, but after the first complete charge it remains above $V_\text{L}$ during subsequent discharges.
    }
    \label{fig:odes}
    \end{figure}

    When the switch in the circuit is closed, the capacitor starts to charge. By virtue of Kirchoff's laws \cite{young2020university, Suits_2020}, one obtains the following system of equations,
    \begin{equation}
    \label{eq:charging_eqs}
    \left\{
    \begin{array}{rl}
           V_\text{S} & = IR_1+ I_\text{L} R_2 + V_\text{L} \\
           V_\text{C} & = I_\text{L} R_2 + V_\text{L}  \\
           I & = I_\text{C} + I_\text{L}
    \end{array}
    \right.,
    \end{equation}
    with the meaning of each variable shown in Fig.~\ref{fig:schemes} (left).      
    Although an exact description of the LED behavior would require accounting for its non-linear nature \cite{Suits_2020}, in Eqn. (\ref{eq:charging_eqs}), we have assumed an ideal behavior with a constant forward voltage drop $V_\mathrm{L}$. As it is later confirmed in Sec.~\ref{sec:validation}, this approximation suffices for the analysis required in this work. Solving this system and using that $I_\text{C} = C dV_\text{C} / dt$, we obtain the following differential equation in $V_\text{C}$
    \begin{equation}\label{eqn:ode_charge}
        CR_1\frac{dV_\text{C}}{dt} + \left(\gamma + 1\right)V_\text{C} - \left(V_\text{S} + \gamma V_\text{L} \right) = 0,
    \end{equation}
    where $\gamma = R_1/R_2$. This is a simple constant coefficient linear ODE, whose analytical solution is
    \begin{equation}\label{eq:vc_charge}
         V_\text{C}(t) = V_{\text{C},\text{max}}\left(1  - \frac{V_{\text{C},\text{max}} - V_\text{C}(0)}{\Vmax}\exp \left[-  t/\tauc\right]\right),
    \end{equation}
    where the time constant of the circuit is $\tauc = CR_1/\left(\gamma +1 \right)$ and $\Vmax = (V_\text{S} + \gamma V_\text{L}) / (\gamma + 1)$ is the capacitor voltage in the \emph{steady-state}. 

     As we will see in the following point, the capacitor voltage starts from $V_\text{L}$ after the first charge-discharge cycle. 
    With this initial condition, $V_\text{C}(0) = V_\text{L}$, we obtain the solution of $V_\text{C}(t)$ during the charge
    \begin{equation}
         V_\text{C}(t) - V_\text{L}
         =
         \frac{V_\text{S} - V_\text{L}}{\gamma + 1}
         \left(1  - \exp \left[-  t/\tauc\right]\right),
    \end{equation}
    which corresponds to the blue curves above the $V_\text{L}$ line in Fig.~\ref{fig:odes}.

    To end this section, some comments are in order.
    First, we can see that, in contrast to the typical $RC$ circuit, the capacitor never reaches the voltage of the source.
    This is caused by the current that powers the LED, which also flows through $R_1$ generating a voltage drop that reduces the voltage available on the capacitor.
    Another difference is that the time constant of the charging circuit is different from the usual $\tau = RC$ due to the way the LED is connected.
    Finally,  if the condition $V_{\text{C}}(0) \ge V_\text{L}$ is not verified (for example in the first cycle), no current flows through $R_2$ and the capacitor and the circuit behave as a usual $RC$ circuit with its corresponding time constant (see first charge in Fig.~\ref{fig:odes}).

\subsection{Discharging circuit}

    We proceed with the discharge process.
    When the switch is opened, the circuit is simplified to a regular $RC$ circuit with an LED in series, which obeys
    \begin{equation}
        V_\text{C} + I^{\prime}R_2 + V_\text{L} = 0.
    \end{equation}
    
    The resultant differential equation is 
    \begin{equation}\label{eqn:ode_discharge}
        CR_2 \dfrac{dV_\text{C}}{dt} + V_\text{C} - V_\text{L} = 0,
    \end{equation}
    which can be solved using the initial condition $V_\text{C}(0) = \Vmax$ to obtain
    \begin{equation}
        V_\text{C}(t) - V_\text{L}
        =
        \left(\Vmax - V_\text{L}\right) \exp\left[-t/\taud\right],
    \end{equation}
    where $\taud = CR_2$ is the new time constant of the circuit.
    This solution is analogous as the one of a regular $RC$ circuit, with $V_\text{C}(t) - V_\text{L}$ replacing $V_\text{C}(t)$.
    As we advanced in the previous section, the capacitor's voltage can not drop below $V_\text{L}$, in contrast to a standard $RC$ circuit where it goes to zero in the $t\to\infty$ limit.
    It is also worth remarking that the time constant in the discharge is different than the one during the charge in the circuit.
    In Section~\ref{sec:altcircuits}, we present an alternative circuit which solves this problem, but also introduces other difficulties.

\subsection{LED intensity and circuit current}
\label{sec:LEDintensitycurrent}

Once the charge and discharge equations are solved, we come back to the main point of this article: measuring the capacitor voltage through the LED brightness.
In order to do so, we derive below the relation between these two quantities.

We start by remembering that in an ideal LED, the brightness, $\Lled$, is proportional to the current $I_\mathrm{L}$ that flows through it.
In our circuit, this is equal to $(V_\text{C} - V_\text{L})/R_2$ by Ohm's law, so that
\begin{equation}\label{eqn:brightness_tension}
    \Lled \propto (V_\mathrm{C} - V_{\mathrm{L}}).
\end{equation}

This equation allows us to determine $\tau$ by measuring the brightness during the charge and discharge of the capacitor.
During the charge, we can use the previous expression to rewrite Eqn.~\eqref{eq:vc_charge} in terms of the LED brightness as
\begin{equation}
\label{eq:brightness_charge}
    \Lled(t)
    =
    \Lledmax\left(1 - \frac{\Lledmax - \Lled(0)}{\Lledmax} \exp[-t/\tauc]\right),
\end{equation}
where $\Lledmax$ is the maximum brightness of the LED, reached when $V_\text{C}(t) = \Vmax$.
From this equation, we can easily measure $\tau$ from the slope of a linear regression of $\log[\Lledmax - \Lled(t)]$ vs $t$, which is given by $-1/\tauc$. In the discharge process, we proceed in an analogous way to obtain in this case
\begin{equation}
\label{eq:brightness_discharge}
    \Lled(t)
    =
    \Lledmax \exp[-t/\taud],
\end{equation}
from where we can obtain $\taud$ from the slope of $\log{\Lled(t)}$ vs. $t$, which is now $-1/\taud$.

\section{The smartphone app}

In order to measure the brightness of the LED, we used the smartphone's light meter using free app \emph{Physics Toolbox Suite} \cite{physicstoolbox}, available for both Android and iOS devices.
This app grants access to a wide range of sensors and measurement tools included in the smartphone, as well as a data logger and graphing tools which can be used to record and analyze data in real-time or to export them in an external file.

\begin{figure}[ht]
    \includegraphics[width=0.65\linewidth]{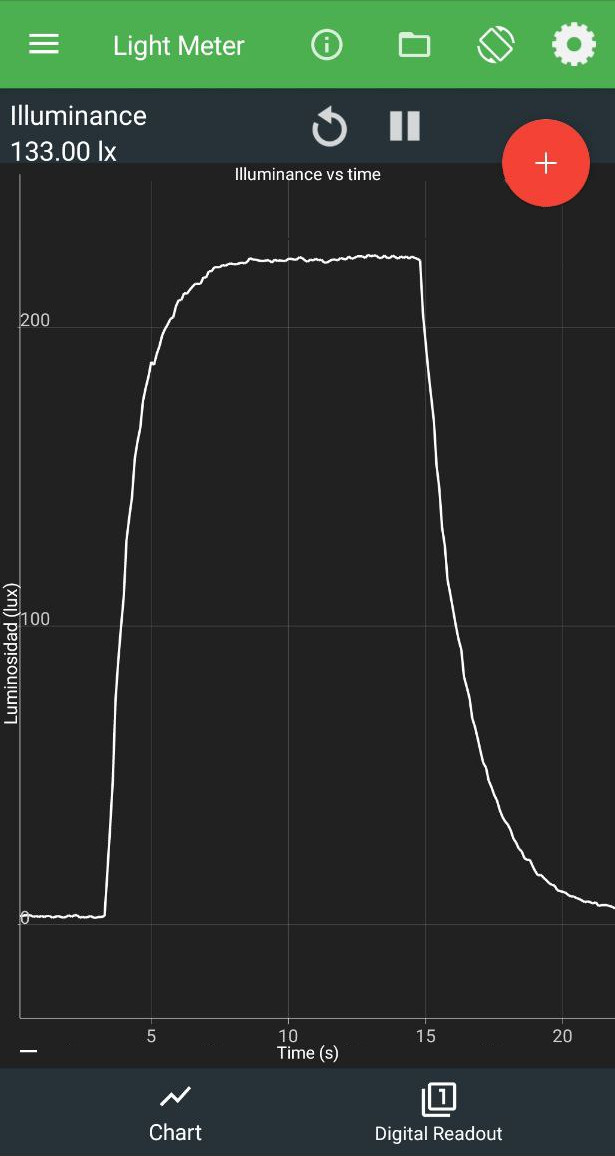}
    \caption{Screenshot of the smartphone app \emph{Physics Toolbox Suite} measuring the LED's brightness during a load-discharge cycle.
    }
    \label{fig:app}
\end{figure}

Specifically, in this experiment we used the light meter of the phone, which can be accessed through ``Light Meter'' button under the main menu.
Fig ~\ref{fig:app} displays a screenshot of the app while measuring the LED's brightness during the experiment.
The main plot shows the illuminance measured by the sensors against time, with the instantaneous illuminance shown at the top left corner in lux.
Another important feature is the red button in the top right corner, which is used to start and stop the data acquisition.
This data is then saved in .csv format for later analysis.

\begin{figure*}[htb]

    \includegraphics[width=0.48\linewidth]{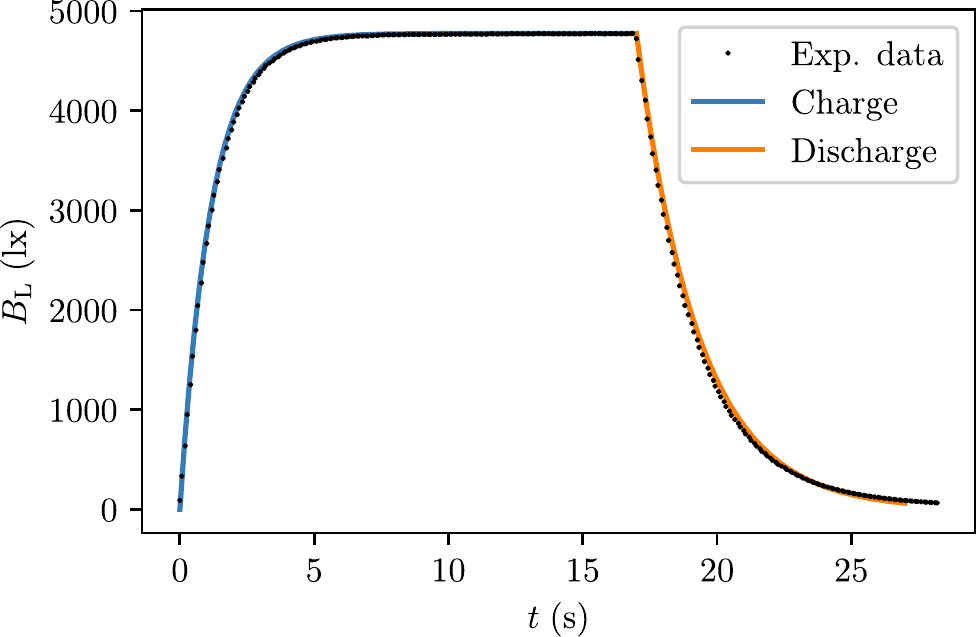}
    \hfill
    \includegraphics[width=0.48\linewidth]{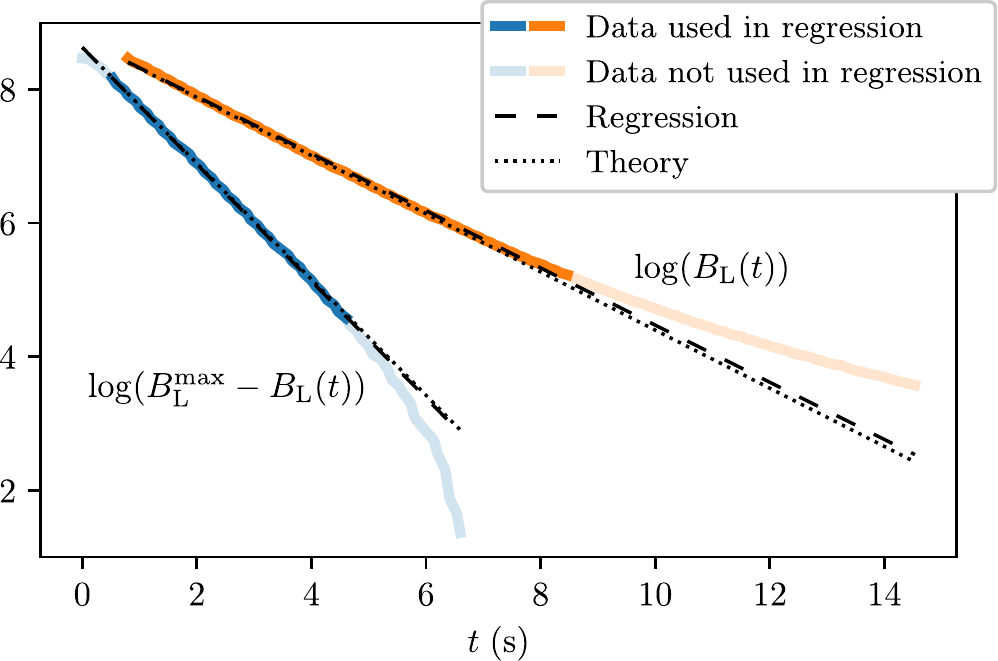}

    \caption{\textbf{Left.} Comparison between the theoretical values obtained by solving the ODEs and the experimental data of the illuminance obtained with the light meter. Both simulations and experiments have been performed for the parameter values given in Table \ref{tab:components}. The agreement of both curves demonstrates the success in measuring the capacitor voltage indirectly through the LED intensity and the proper approach done in the theoretical explanation. \textbf{Right.} Linear regressions of $\log(B_\mathrm{L}^\mathrm{max} - B_\mathrm{L})$ and $\log(B_\mathrm{L})$ during the charging and discharging processes, respectively.
    The slopes of the theoretical and experimental lines are displayed in Table \ref{tab:taus}.}
    \label{fig:experimental}
\end{figure*}

\section{Experimental setup}\label{sec:experimental_setup}

Having introduced the theory in the previous explanations, we proceed to explain the experimental procedure for the experiment. First, we prepared the circuit on a breadboard, as shown in Fig.~\ref{fig:schemes}~(right).
We used an Arduino-like board as a voltage source, but any sort of DC source such as a simple battery or even the power wires of a stripped USB cable would work.
Resistors should be chosen so that the current drawn from the source stays within its specified maximum limit.

Next, we proceed with a crucial step: aligning the LED with the light meter of the phone. This sensor is usually located above the screen, close to the front camera. In order to ensure the proper alignment of the LED, the simplest method is to measure using the app and adjust the LED's position until a maximum in the reading is achieved. Beware that placing the LED too close to the sensor may result in the light intensity exceeding the limits of the sensor, so it is important to check this to avoid saturated readings. A final consideration is the ambient light, which may interfere with the readings of the sensor. Therefore, it is advisable to conduct the experiment in a dimly lit room or to shield the experimental setup with a thick cloth. Fig.~\ref{fig:medidaexp} illustrates a possible arrangement for the experiment. 

Once the setup is prepared, we continue with the measurements. Initially, we perform a full charge and discharge cycle so that in the subsequent charge the capacitor voltage begins at $V_\text{L}$, as explained in Section~\ref{sec:physics}. After this, we initiate the data acquisition through the app and we start the charging process by flipping the switch. The LED's luminosity will gradually increase until the charge is complete.
The end of the charging can be seen in real-time on the phone or can be estimated through the equations. Then, we flip again the switch to measure the brightness during the discharging until the LED becomes completely black.
Finally, we stop the measurement and the data are exported to a .csv file, which can then be analyzed with an external program.

A final remark concerning the light meter lies in the determination of the uncertainty in its measurements. Unfortunately, the accuracy limits of smartphone's light meters are rarely provided by manufacturers. However, it is worth remarking that the experiment aims to offer a valid and accessible way to understand $RC$ circuits. In cases where knowing the measurement uncertainties is mandatory, other well-established methods such as oscilloscope-based techniques are available.

\section{Validation of the method}
\label{sec:validation}

In order to validate the presented method, we performed the experiment using the values of the components displayed in Table \ref{tab:components} and the light meter of a \emph{Xiaomi Redmi Note 8T smartphone}.

\begin{table}[h]
\centering
\renewcommand{\arraystretch}{1.2}
\setlength{\tabcolsep}{12pt}
\begin{tabular}{c|cc}
Component & Value & Error \\
\hline
$C$ & $\SI{490}{\micro\farad}$ & \SI{20}{\micro\farad} \\
$R_1$ & $\SI{4.72}{\kilo\ohm}$ & \SI{0.04}{\kilo\ohm} \\
$R_2$ & $\SI{4.69}{\kilo\ohm}$ & \SI{0.04}{\kilo\ohm} \\
$V_\text{S}$  & $\SI{5.14}{\volt}$ & \SI{0.03}{\volt} \\
$V_\text{L}$  & $\SI{2.50}{\volt}$ & \SI{0.02}{\volt} \\
\end{tabular}
\caption{Components used in the experiment with their corresponding experimental error.}
\label{tab:components}
\end{table}

After following the steps outlined above, we obtained luminosity curves for both the charging and discharging processes, which are shown in Fig.~\ref{fig:experimental}~(left) along with the corresponding expected theoretical curves based on Eqs. \eqref{eq:brightness_charge} and \eqref{eq:brightness_discharge}.
Taking the logarithms in each curve as explained in Section~\ref{sec:LEDintensitycurrent}, we estimated the time constants of both processes using linear regressions (recall that they differ in this circuit).
This is displayed in Fig.~\ref{fig:experimental}~(right), where we can appreciate an outstanding agreement between the theory and the experiment. In particular, we can see that, aside from the end of the curves, the data is linear and that the slopes $m_\text{charge}$ and $m_\text{discharge}$ completely match the theoretical ones $\tauc^{-1}$ and $\taud^{-1}$, respectively. This is further confirmed in Table \ref{tab:taus}, which displays the numerical values of these slopes.
Altogether, these results strongly support the effectiveness of the method presented in this paper.

In reference to the deviations from the theory shown in Fig.~\ref{fig:experimental}~(right), some comments are in order.
First, the difference observed at the end of the $\log(B_\mathrm{L}^\mathrm{max} - B_\mathrm{L})$ curve is probably caused by the finite precision of the light meter, which is unable to properly measure the small differences $B_\mathrm{L}^\mathrm{max} - B_\mathrm{L}$ during the final stages of the charging.
As for the deviations observed at the beginning of the previous curve and at the end of the curve $\log(B_\mathrm{L})$, we suspect that these are caused by a deviation of the LED's behavior from the ideal one at very low currents. In principle, this could be fixed by measuring the $I$ versus $V$ curve of the LED \cite{Suits_2020} and using it to solve the differential equations, but this falls outside the scope of this work.

Finally, it is worth mentioning that the accuracy of the method is quite dependent on the quality of the smartphone sensors. We repeated the experiment with other older smartphones and, while they provided satisfactory estimations with relative errors below $10\%$, they were far from the precision of the results shown above.

\begin{table}[h]
\centering
\renewcommand{\arraystretch}{1.2}
\setlength{\tabcolsep}{12pt}
\begin{tabular}{c|cc}
  & Experimental & Theoretical\\
\hline
$1/\tauc$ & $0.874 \pm 0.003$ & $0.87 \pm 0.04$ \\
$1/\taud$ & $0.4276 \pm 0.0018$ & $0.435 \pm 0.018$ \\
\end{tabular}
\caption{Experimental and theoretical calculations of the time constant in the charge and discharge process. Both values are obtained by performing linear regressions for each set of data: the experimental obtained by the sensor and the theoretical ones given by the solution of the ODEs presented in Section \ref{sec:physics}.}
\label{tab:taus}
\end{table}

\section{Alternative circuit} \label{sec:alternative}
\label{sec:altcircuits}
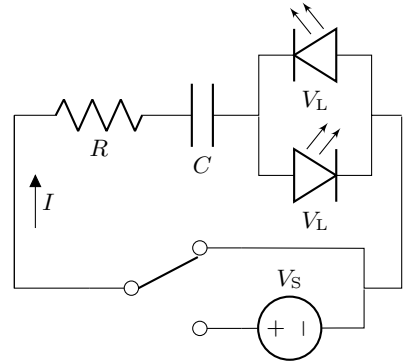
\begin{figure}[b]
\centering
    \begin{circuitikz}[american voltages, every node/.style={inner sep=0,outer sep=0}]
        \draw
        (0,0) [short, -] to ++(0.5,0)
          to [R, l_=$R$] ++(1.25,0)
          to [C, l_=$C$] ++(1.5,0) node(sepled){}
          to [short, -] ++(0,0.8) 
          to [empty led, invert, l_=$V_\mathrm{L}$] ++(1.5,0) 
          to [short, -] ++(0,-0.8) 
    
        (sepled) to [short, -] ++(0,-0.8) 
          to [empty led, l_=$V_\mathrm{L}$] ++(1.5,0) 
          to [short, -] ++(0,+0.8)
          to [short, -] ++(0.4,0) node(topright){}
        
        (0,0) [short,-] to (0,-2.3)
          to [short, -] ++(1.,0) node[spdt, scale=1.7, anchor=in] (Sw) {}

        (topright) to [short, -] ++(0,-2.3) 
          to [short, -] ++(-0.5, 0) node(sepswitch){}
          to [short, -] ++(0, 0.53)
          to [short, -] (Sw.out 1) 
    
        (sepswitch) to [short, -] ++(0, -0.53)
          to [short, -] ++(-0.35, 0)
          to [V, l_=$V_\mathrm{S}$, invert] (Sw.out 2)
          ;

        \draw
        (0,-2.3) [short, f_=$\;I$, -] to (0,0);
    \end{circuitikz}
    
    \caption{Alternative to the original circuit presented in Section \ref{sec:alternative}.}
    \label{fig:altcircuits}
\end{figure}

 In the following discussion, we will explore another possible circuit, depicted in Fig.~\ref{fig:altcircuits}. As we will see, this new circuit solves some of the problems in the original circuit but presents other disadvantages.
 
This proposed layout is just a slight modification of the standard $RC$ circuit with two parallel LEDs in opposite directions, which are placed in series with the rest of the components. This new arrangement with two diodes is required because the current flows in different directions depending on the state of the switch. 
One of the advantages of this configuration is that its equations are as straightforward to solve as those of the standard $RC$ circuit. In particular, it can be easily shown that after the first cycle, the voltage at the capacitor and the current across the correctly oriented LED (and therefore its brightness) are given by the following expressions:
\begin{itemize}
    
    \item Charge:
    \begin{align}
        V_{\text{C}}(t) &= V_\text{L} + (V_\text{S} - 2V_\text{L})(1 - \exp \left[-  t/RC\right]).
        \\
        I(t) &= \frac{V_\text{S} - 2V_\text{L}}{R}\exp \left[-  t/RC\right].
    \end{align}
    
    \item Discharge:
    \begin{align}
        V_{\text{C}}(t) - V_\text{L} &= (V_\text{S} - 2V_\text{L})\exp \left[-  t/RC\right].
        \\
        I(t) &= -\frac{V_\text{S} - 2V_\text{L}}{R}\exp \left[-  t/RC\right].
    \end{align}    
    \end{itemize}
In order to solve these ODEs, we have considered the initial conditions $V_\text{C}(0) = V_\text{L}$ for the charge and $V_\text{C}(0) = V_\text{S} - V_\text{L}$ for the discharge.

Unlike the circuit presented in Section \ref{sec:physics}, in this case, both the charge and discharge processes follow the usual time constant of the $RC$ circuit, $\tau=RC$. However, the current $I$ across the LED in this circuit is proportional to the voltage drop in the resistor, and therefore its brightness displays an exponential decay in both processes. This makes the relation between the capacitor state and the LED's brightness less intuitive, rendering this arrangement less suitable for an illustrative demonstration of the $RC$ circuit.

\section{Conclusions}
In this work, we have proposed a novel and effective approach for measuring capacitor charge and discharge processes using a LED and the light meter of a smartphone.
By exploiting the LED's light intensity as an indicator of the voltage across the capacitor, this technique provides an accurate and reliable measurement of the capacitor state during the charge and discharge cycles. 
One of the main advantages of this technique lies in its simplicity and ease of implementation, as it only requires widely available and basic equipment.
As a consequence, it can be readily used in educational settings to teach fundamental concepts of electricity and electronics. 

This work opens the door to developing other experiments in electronics following the same concept, where the use of smartphones could be integrated to provide real-time monitoring of the magnitudes of interest.
Moreover, future work in this area might focus on exploring new applications in
different fields, where the magnitudes of interest can be indirectly assessed by monitoring the brightness of LEDs.
Overall, the use of smartphones represents a promising avenue for advancing in the field of electronics and physics education. Its potential applications are numerous and diverse, and its simplicity and accuracy make it an attractive option for researchers and teachers alike. 
Its integration into physics education will certainly lead us to improved and innovative teaching methodologies.

\section*{Acknowledgements}
\noindent The research leading to these results has received funding from the fellowship FPU17/02191 and FPU20/02835, projects Ref. PID2020-113681GB-I00, Ref. PID2021-128970OA-I0 from the Spanish Ministry and Agencia Estatal de Investigación (AEI) and from FEDER/Junta de Andalucía programs A.FQM.752.UGR20 and P20\_00173. The authors would like to thank F. de los Santos and A. Ortiz-Mora for their advice and insightful comments.

\section*{Data Availanility Statement}
The data that support the findings of this study are available upon reasonable request from the
authors.\url{ https://doi.org/10.5281/zenodo.7994343}.

\bibliography{biblio}

\begin{thebibliography}{28}%
\makeatletter
\providecommand \@ifxundefined [1]{%
 \@ifx{#1\undefined}
}%
\providecommand \@ifnum [1]{%
 \ifnum #1\expandafter \@firstoftwo
 \else \expandafter \@secondoftwo
 \fi
}%
\providecommand \@ifx [1]{%
 \ifx #1\expandafter \@firstoftwo
 \else \expandafter \@secondoftwo
 \fi
}%
\providecommand \natexlab [1]{#1}%
\providecommand \enquote  [1]{``#1''}%
\providecommand \bibnamefont  [1]{#1}%
\providecommand \bibfnamefont [1]{#1}%
\providecommand \citenamefont [1]{#1}%
\providecommand \href@noop [0]{\@secondoftwo}%
\providecommand \href [0]{\begingroup \@sanitize@url \@href}%
\providecommand \@href[1]{\@@startlink{#1}\@@href}%
\providecommand \@@href[1]{\endgroup#1\@@endlink}%
\providecommand \@sanitize@url [0]{\catcode `\\12\catcode `\$12\catcode
  `\&12\catcode `\#12\catcode `\^12\catcode `\_12\catcode `\%12\relax}%
\providecommand \@@startlink[1]{}%
\providecommand \@@endlink[0]{}%
\providecommand \url  [0]{\begingroup\@sanitize@url \@url }%
\providecommand \@url [1]{\endgroup\@href {#1}{\urlprefix }}%
\providecommand \urlprefix  [0]{URL }%
\providecommand \Eprint [0]{\href }%
\providecommand \doibase [0]{http://dx.doi.org/}%
\providecommand \selectlanguage [0]{\@gobble}%
\providecommand \bibinfo  [0]{\@secondoftwo}%
\providecommand \bibfield  [0]{\@secondoftwo}%
\providecommand \translation [1]{[#1]}%
\providecommand \BibitemOpen [0]{}%
\providecommand \bibitemStop [0]{}%
\providecommand \bibitemNoStop [0]{.\EOS\space}%
\providecommand \EOS [0]{\spacefactor3000\relax}%
\providecommand \BibitemShut  [1]{\csname bibitem#1\endcsname}%
\let\auto@bib@innerbib\@empty
\bibitem [{\citenamefont {La~Cour}\ \emph {et~al.}(2022)\citenamefont
  {La~Cour}, \citenamefont {Maynard}, \citenamefont {Shroff}, \citenamefont
  {Ko},\ and\ \citenamefont {Ellis}}]{9951268}%
  \BibitemOpen
  \bibfield  {author} {\bibinfo {author} {\bibfnamefont {Brian~R.}\
  \bibnamefont {La~Cour}}, \bibinfo {author} {\bibfnamefont {Maria}\
  \bibnamefont {Maynard}}, \bibinfo {author} {\bibfnamefont {Parth}\
  \bibnamefont {Shroff}}, \bibinfo {author} {\bibfnamefont {Gabriel}\
  \bibnamefont {Ko}}, \ and\ \bibinfo {author} {\bibfnamefont {Evan}\
  \bibnamefont {Ellis}},\ }\bibfield  {title} {\enquote {\bibinfo {title} {{The
  Virtual Quantum Optics Laboratory}},}\ }in\ \href {\doibase
  10.1109/QCE53715.2022.00091} {\emph {\bibinfo {booktitle} {2022 IEEE
  International Conference on Quantum Computing and Engineering (QCE)}}}\
  (\bibinfo {year} {2022})\ pp.\ \bibinfo {pages} {677--687}\BibitemShut
  {NoStop}%
\bibitem [{\citenamefont {Monteiro}\ \emph {et~al.}(2022)\citenamefont
  {Monteiro}, \citenamefont {Stari},\ and\ \citenamefont
  {Martí}}]{Monteiro_2023}%
  \BibitemOpen
  \bibfield  {author} {\bibinfo {author} {\bibfnamefont {Martín}\ \bibnamefont
  {Monteiro}}, \bibinfo {author} {\bibfnamefont {Cecilia}\ \bibnamefont
  {Stari}}, \ and\ \bibinfo {author} {\bibfnamefont {Arturo~C}\ \bibnamefont
  {Martí}},\ }\bibfield  {title} {\enquote {\bibinfo {title} {{A home-lab
  experiment: resonance and sound speed using telescopic vacuum cleaner
  pipes}},}\ }\href {\doibase 10.1088/1361-6552/ac9ae1} {\bibfield  {journal}
  {\bibinfo  {journal} {Physics Education}\ }\textbf {\bibinfo {volume} {58}},\
  \bibinfo {pages} {013003} (\bibinfo {year} {2022})}\BibitemShut {NoStop}%
\bibitem [{\citenamefont {Haldolaarachchige}\ and\ \citenamefont
  {Hettiarachchilage}(2020{\natexlab{a}})}]{haldolaarachchige2020lab}%
  \BibitemOpen
  \bibfield  {author} {\bibinfo {author} {\bibfnamefont {N.}~\bibnamefont
  {Haldolaarachchige}}\ and\ \bibinfo {author} {\bibfnamefont {K.}~\bibnamefont
  {Hettiarachchilage}},\ }\href@noop {} {\enquote {\bibinfo {title} {{Lab
  Manual of Introductory Physics-I for Virtual Teaching}},}\ } (\bibinfo {year}
  {2020}{\natexlab{a}}),\ \Eprint {http://arxiv.org/abs/2012.09151}
  {arXiv:2012.09151 [physics.ed-ph]} \BibitemShut {NoStop}%
\bibitem [{\citenamefont {Haldolaarachchige}\ and\ \citenamefont
  {Hettiarachchilage}(2020{\natexlab{b}})}]{haldolaarachchige2020introductory}%
  \BibitemOpen
  \bibfield  {author} {\bibinfo {author} {\bibfnamefont {N.}~\bibnamefont
  {Haldolaarachchige}}\ and\ \bibinfo {author} {\bibfnamefont {K.}~\bibnamefont
  {Hettiarachchilage}},\ }\href@noop {} {\enquote {\bibinfo {title}
  {{Introductory {EM} Lab Manual for Virtual Teaching}},}\ } (\bibinfo {year}
  {2020}{\natexlab{b}}),\ \Eprint {http://arxiv.org/abs/2012.13278}
  {arXiv:2012.13278 [physics.ed-ph]} \BibitemShut {NoStop}%
\bibitem [{\citenamefont {Haldolaarachchige}\ and\ \citenamefont
  {Hettiarachchilage}(2021)}]{haldolaarachchige2021set}%
  \BibitemOpen
  \bibfield  {author} {\bibinfo {author} {\bibfnamefont {N.}~\bibnamefont
  {Haldolaarachchige}}\ and\ \bibinfo {author} {\bibfnamefont {K.}~\bibnamefont
  {Hettiarachchilage}},\ }\href@noop {} {\enquote {\bibinfo {title} {{A Set of
  Virtual Experiments of Fluids, Waves, Thermodynamics, Optics, and Modern
  Physics for Virtual Teaching of Introductory Physics}},}\ } (\bibinfo {year}
  {2021}),\ \Eprint {http://arxiv.org/abs/2101.00993} {arXiv:2101.00993
  [physics.ed-ph]} \BibitemShut {NoStop}%
\bibitem [{\citenamefont {Raman}\ \emph {et~al.}(2022)\citenamefont {Raman},
  \citenamefont {Achuthan}, \citenamefont {Nair},\ and\ \citenamefont
  {Nedungadi}}]{Raman_Achuthan_Nair_Nedungadi_2022}%
  \BibitemOpen
  \bibfield  {author} {\bibinfo {author} {\bibfnamefont {R.}~\bibnamefont
  {Raman}}, \bibinfo {author} {\bibfnamefont {K.}~\bibnamefont {Achuthan}},
  \bibinfo {author} {\bibfnamefont {Vinith~K.}\ \bibnamefont {Nair}}, \ and\
  \bibinfo {author} {\bibfnamefont {P.}~\bibnamefont {Nedungadi}},\ }\bibfield
  {title} {\enquote {\bibinfo {title} {{Virtual Laboratories - A historical
  review and bibliometric analysis of the past three decades}},}\ }\href
  {\doibase 10.1007/s10639-022-11058-9} {\bibfield  {journal} {\bibinfo
  {journal} {Education and Information Technologies}\ }\textbf {\bibinfo
  {volume} {27}},\ \bibinfo {pages} {11055–11087} (\bibinfo {year}
  {2022})}\BibitemShut {NoStop}%
\bibitem [{\citenamefont {Abdulwahed}\ and\ \citenamefont
  {Nagy}(2014)}]{https://doi.org/10.1002/cae.20536}%
  \BibitemOpen
  \bibfield  {author} {\bibinfo {author} {\bibfnamefont {M.}~\bibnamefont
  {Abdulwahed}}\ and\ \bibinfo {author} {\bibfnamefont {Z.~K.}\ \bibnamefont
  {Nagy}},\ }\bibfield  {title} {\enquote {\bibinfo {title} {{The impact of
  different preparation modes on enhancing the undergraduate process control
  engineering laboratory: A comparative study}},}\ }\href {\doibase
  https://doi.org/10.1002/cae.20536} {\bibfield  {journal} {\bibinfo  {journal}
  {Computer Applications in Engineering Education}\ }\textbf {\bibinfo {volume}
  {22}},\ \bibinfo {pages} {110--119} (\bibinfo {year} {2014})}\BibitemShut
  {NoStop}%
\bibitem [{\citenamefont {Gunawan}\ \emph {et~al.}(2018)\citenamefont
  {Gunawan}, \citenamefont {Nisrina}, \citenamefont {Suranti}, \citenamefont
  {Herayanti},\ and\ \citenamefont {Rahmatiah}}]{Gunawan_2018}%
  \BibitemOpen
  \bibfield  {author} {\bibinfo {author} {\bibfnamefont {G.}~\bibnamefont
  {Gunawan}}, \bibinfo {author} {\bibfnamefont {N.}~\bibnamefont {Nisrina}},
  \bibinfo {author} {\bibfnamefont {N.~M.~Y.}\ \bibnamefont {Suranti}},
  \bibinfo {author} {\bibfnamefont {L.}~\bibnamefont {Herayanti}}, \ and\
  \bibinfo {author} {\bibfnamefont {R}~\bibnamefont {Rahmatiah}},\ }\bibfield
  {title} {\enquote {\bibinfo {title} {{Virtual Laboratory to Improve
  Students’ Conceptual Understanding in Physics Learning}},}\ }\href
  {\doibase 10.1088/1742-6596/1108/1/012049} {\bibfield  {journal} {\bibinfo
  {journal} {Journal of Physics: Conference Series}\ }\textbf {\bibinfo
  {volume} {1108}},\ \bibinfo {pages} {012049} (\bibinfo {year}
  {2018})}\BibitemShut {NoStop}%
\bibitem [{\citenamefont {Casaburo}(2021)}]{casaburo2021teaching}%
  \BibitemOpen
  \bibfield  {author} {\bibinfo {author} {\bibfnamefont {Fausto}\ \bibnamefont
  {Casaburo}},\ }\bibfield  {title} {\enquote {\bibinfo {title} {{Teaching
  physics by Arduino during COVID-19 pandemic: the free falling body
  experiment}},}\ }\href {\doibase 10.1088/1361-6552/ac1b39} {\bibfield
  {journal} {\bibinfo  {journal} {Physics Education}\ }\textbf {\bibinfo
  {volume} {56}},\ \bibinfo {pages} {063001} (\bibinfo {year}
  {2021})}\BibitemShut {NoStop}%
\bibitem [{\citenamefont {Franco}\ \emph {et~al.}(2023)\citenamefont {Franco},
  \citenamefont {Serbena}, \citenamefont {Mattoso}, \citenamefont
  {de~Oliveira~Wiener}, \citenamefont {Yokaichiya},\ and\ \citenamefont
  {Fujimoto}}]{Franco_2023}%
  \BibitemOpen
  \bibfield  {author} {\bibinfo {author} {\bibfnamefont {M.~K. K.~D.}\
  \bibnamefont {Franco}}, \bibinfo {author} {\bibfnamefont {J.~P.~M.}\
  \bibnamefont {Serbena}}, \bibinfo {author} {\bibfnamefont {N.}~\bibnamefont
  {Mattoso}}, \bibinfo {author} {\bibfnamefont {D.}~\bibnamefont
  {de~Oliveira~Wiener}}, \bibinfo {author} {\bibfnamefont {F.}~\bibnamefont
  {Yokaichiya}}, \ and\ \bibinfo {author} {\bibfnamefont {M.~M.}\ \bibnamefont
  {Fujimoto}},\ }\bibfield  {title} {\enquote {\bibinfo {title} {{Feasibility
  of teaching experimental physics during COVID-19 pandemic}},}\ }\href
  {\doibase 10.1088/1361-6552/acbf1d} {\bibfield  {journal} {\bibinfo
  {journal} {Physics Education}\ }\textbf {\bibinfo {volume} {58}},\ \bibinfo
  {pages} {035030} (\bibinfo {year} {2023})}\BibitemShut {NoStop}%
\bibitem [{\citenamefont {Bjurholt}\ and\ \citenamefont
  {Bøe}(2022)}]{Bjurholt_2023}%
  \BibitemOpen
  \bibfield  {author} {\bibinfo {author} {\bibfnamefont {Nikolai}\ \bibnamefont
  {Bjurholt}}\ and\ \bibinfo {author} {\bibfnamefont {Maria~Vetleseter}\
  \bibnamefont {Bøe}},\ }\bibfield  {title} {\enquote {\bibinfo {title}
  {{Remote physics teaching during the COVID-19 pandemic: losses and potential
  gains}},}\ }\href {\doibase 10.1088/1361-6552/ac96be} {\bibfield  {journal}
  {\bibinfo  {journal} {Physics Education}\ }\textbf {\bibinfo {volume} {58}},\
  \bibinfo {pages} {015004} (\bibinfo {year} {2022})}\BibitemShut {NoStop}%
\bibitem [{\citenamefont {Guo}(2020)}]{Guo_2020}%
  \BibitemOpen
  \bibfield  {author} {\bibinfo {author} {\bibfnamefont {Siming}\ \bibnamefont
  {Guo}},\ }\bibfield  {title} {\enquote {\bibinfo {title} {{Synchronous versus
  asynchronous online teaching of physics during the COVID-19 pandemic}},}\
  }\href {\doibase 10.1088/1361-6552/aba1c5} {\bibfield  {journal} {\bibinfo
  {journal} {Physics Education}\ }\textbf {\bibinfo {volume} {55}},\ \bibinfo
  {pages} {065007} (\bibinfo {year} {2020})}\BibitemShut {NoStop}%
\bibitem [{\citenamefont {{University of Colorado}}()}]{phet}%
  \BibitemOpen
  \bibfield  {author} {\bibinfo {author} {\bibnamefont {{University of
  Colorado}}},\ }\href@noop {} {\enquote {\bibinfo {title} {{PhET: Free online
  Physics, Chemistry, Biology, Earth Science and Math simulations}},}\
  }\bibinfo {howpublished} {\url{https://phet.colorado.edu/}}\BibitemShut
  {NoStop}%
\bibitem [{\citenamefont {{University of St Andrews}}()}]{quvis}%
  \BibitemOpen
  \bibfield  {author} {\bibinfo {author} {\bibnamefont {{University of St
  Andrews}}},\ }\href@noop {} {\enquote {\bibinfo {title} {{QuVis: The Quantum
  Mechanics Visualisation Project}},}\ }\bibinfo {howpublished}
  {\url{https://www.st-andrews.ac.uk/physics/quvis/}}\BibitemShut {NoStop}%
\bibitem [{ard()}]{arduino}%
  \BibitemOpen
  \href@noop {} {\enquote {\bibinfo {title} {{Arduino}},}\ }\bibinfo
  {howpublished} {\url{https://www.arduino.cc/}}\BibitemShut {NoStop}%
\bibitem [{\citenamefont {Bouquet}\ \emph {et~al.}(2017)\citenamefont
  {Bouquet}, \citenamefont {Bobroff}, \citenamefont {Fuchs-Gallezot},\ and\
  \citenamefont {Maurines}}]{bouquet2017project}%
  \BibitemOpen
  \bibfield  {author} {\bibinfo {author} {\bibfnamefont {F.}~\bibnamefont
  {Bouquet}}, \bibinfo {author} {\bibfnamefont {J.}~\bibnamefont {Bobroff}},
  \bibinfo {author} {\bibfnamefont {M.}~\bibnamefont {Fuchs-Gallezot}}, \ and\
  \bibinfo {author} {\bibfnamefont {L.}~\bibnamefont {Maurines}},\ }\bibfield
  {title} {\enquote {\bibinfo {title} {{Project-based physics labs using
  low-cost open-source hardware}},}\ }\href {\doibase 10.1119/1.4972043}
  {\bibfield  {journal} {\bibinfo  {journal} {American Journal of Physics}\
  }\textbf {\bibinfo {volume} {85}},\ \bibinfo {pages} {216--222} (\bibinfo
  {year} {2017})}\BibitemShut {NoStop}%
\bibitem [{\citenamefont {Haugen}\ and\ \citenamefont
  {Moore}(2014)}]{haugen2014model}%
  \BibitemOpen
  \bibfield  {author} {\bibinfo {author} {\bibfnamefont {Andrew~J.}\
  \bibnamefont {Haugen}}\ and\ \bibinfo {author} {\bibfnamefont {Nathan~T.}\
  \bibnamefont {Moore}},\ }\href@noop {} {\enquote {\bibinfo {title} {{A model
  for including Arduino microcontroller programming in the introductory physics
  lab}},}\ } (\bibinfo {year} {2014}),\ \Eprint
  {http://arxiv.org/abs/1407.7613} {arXiv:1407.7613 [physics.ed-ph]}
  \BibitemShut {NoStop}%
\bibitem [{\citenamefont {Galeriu}\ \emph {et~al.}(2015)\citenamefont
  {Galeriu}, \citenamefont {Letson},\ and\ \citenamefont
  {Esper}}]{galeriu2015arduino}%
  \BibitemOpen
  \bibfield  {author} {\bibinfo {author} {\bibfnamefont {Calin}\ \bibnamefont
  {Galeriu}}, \bibinfo {author} {\bibfnamefont {Cheryl}\ \bibnamefont
  {Letson}}, \ and\ \bibinfo {author} {\bibfnamefont {Geoffrey}\ \bibnamefont
  {Esper}},\ }\bibfield  {title} {\enquote {\bibinfo {title} {{An Arduino
  Investigation of the RC Circuit}},}\ }\href {\doibase 10.1119/1.4917435}
  {\bibfield  {journal} {\bibinfo  {journal} {The Physics Teacher}\ }\textbf
  {\bibinfo {volume} {53}},\ \bibinfo {pages} {285--288} (\bibinfo {year}
  {2015})}\BibitemShut {NoStop}%
\bibitem [{\citenamefont {Erol}\ and\ \citenamefont {Oğur}(2023)}]{Erol_2023}%
  \BibitemOpen
  \bibfield  {author} {\bibinfo {author} {\bibfnamefont {Mustafa}\ \bibnamefont
  {Erol}}\ and\ \bibinfo {author} {\bibfnamefont {Mehmet}\ \bibnamefont
  {Oğur}},\ }\bibfield  {title} {\enquote {\bibinfo {title} {{Teaching large
  angle pendulum via Arduino based STEM education material}},}\ }\href
  {\doibase 10.1088/1361-6552/accef4} {\bibfield  {journal} {\bibinfo
  {journal} {Physics Education}\ }\textbf {\bibinfo {volume} {58}},\ \bibinfo
  {pages} {045001} (\bibinfo {year} {2023})}\BibitemShut {NoStop}%
\bibitem [{\citenamefont {Kuhn}\ and\ \citenamefont
  {Vogt}(2022)}]{kuhn2022smartphones}%
  \BibitemOpen
  \bibfield  {author} {\bibinfo {author} {\bibfnamefont {Jochen}\ \bibnamefont
  {Kuhn}}\ and\ \bibinfo {author} {\bibfnamefont {Patrik}\ \bibnamefont
  {Vogt}},\ }\href {\doibase 10.1007/978-3-030-94044-7} {\emph {\bibinfo
  {title} {{Smartphones as Mobile Minilabs in Physics}}}}\ (\bibinfo
  {publisher} {Springer International Publishing},\ \bibinfo {year}
  {2022})\BibitemShut {NoStop}%
\bibitem [{\citenamefont {Imtinan}\ and\ \citenamefont
  {Kuswanto}(2023)}]{JIPF4167}%
  \BibitemOpen
  \bibfield  {author} {\bibinfo {author} {\bibfnamefont {N.}~\bibnamefont
  {Imtinan}}\ and\ \bibinfo {author} {\bibfnamefont {H.}~\bibnamefont
  {Kuswanto}},\ }\bibfield  {title} {\enquote {\bibinfo {title} {{The Use of
  Phyphox Application in Physics Experiments: A Literature Review}},}\ }\href
  {\doibase 10.26737/jipf.v8i2.4167} {\bibfield  {journal} {\bibinfo  {journal}
  {JIPF (Jurnal Ilmu Pendidikan Fisika)}\ }\textbf {\bibinfo {volume} {8}},\
  \bibinfo {pages} {183--191} (\bibinfo {year} {2023})}\BibitemShut {NoStop}%
\bibitem [{\citenamefont {Gillen}\ and\ \citenamefont
  {Gillen}(2022)}]{gillen2022magnetic}%
  \BibitemOpen
  \bibfield  {author} {\bibinfo {author} {\bibfnamefont {G.~D.}\ \bibnamefont
  {Gillen}}\ and\ \bibinfo {author} {\bibfnamefont {K.}~\bibnamefont
  {Gillen}},\ }\href@noop {} {\enquote {\bibinfo {title} {{Magnetic Field
  Experiments Using a Phone}},}\ } (\bibinfo {year} {2022}),\ \Eprint
  {http://arxiv.org/abs/2212.02258} {arXiv:2212.02258 [physics.ed-ph]}
  \BibitemShut {NoStop}%
\bibitem [{\citenamefont {Gastaldi}\ and\ \citenamefont
  {Campardo}(2020)}]{gastaldi2020electronic}%
  \BibitemOpen
  \bibfield  {author} {\bibinfo {author} {\bibfnamefont {R.}~\bibnamefont
  {Gastaldi}}\ and\ \bibinfo {author} {\bibfnamefont {G.}~\bibnamefont
  {Campardo}},\ }\href@noop {} {\emph {\bibinfo {title} {Electronic experiences
  in a virtual lab}}}\ (\bibinfo  {publisher} {Springer},\ \bibinfo {year}
  {2020})\BibitemShut {NoStop}%
\bibitem [{\citenamefont {Organtini}(2021)}]{organtini2021physics}%
  \BibitemOpen
  \bibfield  {author} {\bibinfo {author} {\bibfnamefont {Giovanni}\
  \bibnamefont {Organtini}},\ }\href@noop {} {\emph {\bibinfo {title} {{Physics
  Experiments with Arduino and Smartphones}}}}\ (\bibinfo  {publisher}
  {Springer},\ \bibinfo {year} {2021})\BibitemShut {NoStop}%
\bibitem [{\citenamefont {University}()}]{phyphox}%
  \BibitemOpen
  \bibfield  {author} {\bibinfo {author} {\bibfnamefont {RWTH~Aachen}\
  \bibnamefont {University}},\ }\href@noop {} {\enquote {\bibinfo {title}
  {{Phyphox [mobile app]}},}\ }\bibinfo {howpublished}
  {\url{https://phyphox.org/}}\BibitemShut {NoStop}%
\bibitem [{\citenamefont {{Vieyra Software}}()}]{physicstoolbox}%
  \BibitemOpen
  \bibfield  {author} {\bibinfo {author} {\bibnamefont {{Vieyra Software}}},\
  }\href@noop {} {\enquote {\bibinfo {title} {{Physics Toolbox Sensor Suite
  [mobile app]}},}\ }\bibinfo {howpublished}
  {\url{https://www.vieyrasoftware.net/physics-toolbox-sensor-suite}}\BibitemShut
  {NoStop}%
\bibitem [{\citenamefont {Young}\ and\ \citenamefont
  {Freedman}(2020)}]{young2020university}%
  \BibitemOpen
  \bibfield  {author} {\bibinfo {author} {\bibfnamefont {H.D.}\ \bibnamefont
  {Young}}\ and\ \bibinfo {author} {\bibfnamefont {R.A.}\ \bibnamefont
  {Freedman}},\ }\href@noop {} {\emph {\bibinfo {title} {{University
  Physics}}}}\ (\bibinfo  {publisher} {Pearson},\ \bibinfo {year}
  {2020})\BibitemShut {NoStop}%
\bibitem [{\citenamefont {Suits}(2020)}]{Suits_2020}%
  \BibitemOpen
  \bibfield  {author} {\bibinfo {author} {\bibfnamefont {B.~H.}\ \bibnamefont
  {Suits}},\ }\href {\doibase 10.1007/978-3-030-39088-4} {\emph {\bibinfo
  {title} {{Electronics for Physicists: An Introduction}}}},\ Undergraduate
  Lecture Notes in Physics\ (\bibinfo  {publisher} {Springer International
  Publishing},\ \bibinfo {year} {2020})\BibitemShut {NoStop}%
\end{thebibliography}%

\end{document}